# Formation Mechanism of Guided Resonances and Bound States in the Continuum in Photonic Crystal Slabs

Xingwei Gao[1†], Chia Wei Hsu[2,3*], Bo Zhen[2,4], Xiao Lin[1,2], John D. Joannopoulos[2], Marin Soljačić[2] and Hongsheng Chen[1*]

[1]The Electromagnetics Academy at Zhejiang University, Zhejiang University, Hangzhou 310027, China

[2]Research Laboratory of Electronics, Massachusetts Institute of Technology, Cambridge, Massachusetts 02139, USA

[3]Department of Applied Physics, Yale University, New Haven, Connecticut 06520, USA

[4]Physics Department and Solid State Institute, Technion, Haifa 32000, Israel.

*Correspondence: Chia Wei Hsu, Email: chiawei.hsu@yale.edu; or Hongsheng Chen, Email: hansomchen@zju.edu.cn

**We develop a formalism, based on the mode expansion method, to describe the guided resonances and bound states in the continuum (BICs) in photonic crystal slabs with one-dimensional periodicity. This approach provides analytic insights to the formation mechanisms of these states: the guided resonances arise from the transverse Fabry-Pérot condition, and the divergence of the resonance lifetimes at the BICs is explained by a destructive interference of radiation from different propagating components inside the slab. We show BICs at the center and on the edge of the Brillouin zone protected by symmetry, as well as BICs at generic wave vectors not protected by symmetry.**

## Introduction

Conventionally, the confinement of waves is achieved by spectrally separating the bound state away from the continuum of radiating waves that can carry energy away—examples include electronic bound states at negative energies and light guided below the light line or inside a photonic bandgap. Bound states in the continuum (BICs) are special states that remain localized and have infinite lifetimes even though they reside inside the continuum[1]. Historically, BIC was first proposed by von Neumann and Wigner for an electron in an engineered potential[2], although such an electron BIC has never been achieved. More recently, optical BICs have been experimentally realized in a range of photonic systems[3-9], although the definition of BICs in periodic structures is not identical to the original definition of BIC in an isolated structure[10]. In periodic structures, BICs may be found by varying the incident angle without tuning the structure, which makes their realization relatively simple[3,5,6,11-22]. Photonic crystal (PhC) slabs—dielectric slabs with periodically modulated refractive index[11,23,24]—are particularly attractive given their macroscopic sizes and ease of fabrication. The guided resonances and BICs in PhC slabs have been used for a wide range of applications from lasers[25-19] to sensors[13,30,31] and filters[32]. The wave confinement mechanisms of BICs are distinct from those of conventional bound states. In the presence of multiple resonances, BICs can arise from the destructive interference of radiation from the different resonances[13-16,33,34]. However, when only one resonance is present, the theoretical studies of BICs rely on numerical simulations that are time consuming and provide little insight to the formation mechanism. Here, we develop a mode expansion method that is capable of describing guided resonances and BICs in slabs quantitatively without any other approximations except for a truncation of the basis size, and yet it also provides the fast speed and insights of an analytic model. In this formalism, guided resonances require the round-trip transverse phase shift for each in-slab propagating mode to be an integer multiple of $2\pi$, and BICs can be understood as arising from the destructive interference of radiation from different propagating waves inside the slab. We also apply this approach to predict optical BICs in periodic inhomogeneous backgrounds, where BICs can be found both at the center and on the edge of the Brillouin zone due to symmetry protection, as well as inside the Brillouin zone where they are not protected by symmetry. Although we only show examples for systems with one dimensional periodicity, our mode-expansion method can be easily extended into systems that are periodic in two dimensions.

## Materials and Methods

### Mode expansion method

For simplicity, we consider TM modes (with electric fields $\mathbf{E} = E_z\hat{z}$) in a PhC slab that is periodic in $y$ and uniform in $z$ (Fig. 1a); the generalization to TE modes and to PhC slabs with two-dimensional periodicity is straightforward. We consider structures that are mirror symmetric in the normal direction $x$ (this symmetry is necessary to reduce the number of radiation channels[6]) and where the slab permittivity $\varepsilon_1(y)$ does not vary in the normal direction (common in most fabricated structures of PhC slabs). Here we consider two cases: the permittivity of the surrounding medium $\varepsilon_2(y)$ is a constant (Fig. 1b) or is periodic with the same period $a$ as the slab (Fig. 1c). Any field profile satisfying the wave equation $[\nabla^2 + \varepsilon(x,y)k_0^2]E_z = 0$ can always be expanded in the basis of the eigenmodes of $\varepsilon_1(y)$ and $\varepsilon_2(y)$ -where $k_0 = \omega/c$, is the frequency, and $c$ is the vacuum speed of light. For an even-in-$x$ TM mode with wave vector $k_y$, we can write

$$E_z(x,y) = e^{ik_y y} \begin{cases} \sum_m C_m \dfrac{\cos(\beta_m x)}{\cos(0.5\beta_m h)} u_m(y), & 0 \le x < 0.5h, \\ \sum_m T_m e^{i\gamma_m (x-0.5h)} \vartheta_m(y), & x > 0.5h, \end{cases} \quad (1)$$

where $h$ is the slab thickness. Here, $u_m(y)$, $\vartheta_m(y)$, $\beta_m$ and $\gamma_m$ are eigenfunctions and propagation constants inside and outside the slab:

$$\widehat{H}_1 u_m(y) = \beta_m^2 u_m(y), \qquad (2.1)$$

$$\widehat{H}_2 \vartheta_m(y) = \gamma_m^2 \vartheta_m(y), \qquad (2.2)$$

with $\widehat{H}_l = (\partial/\partial y + ik_y)^2 + k_0^2 \varepsilon_l(y)$ for $l = 1,2$. $\widehat{H}_1$ and $\widehat{H}_2$ are Hermitian operators subject to periodic boundary condition $u_m(y+a) = u_m(y)$, $\vartheta_m(y+a) = \vartheta_m(y)$. For a given frequency and $k_y$, there will be a finite number of eigenmodes with $\beta_m^2 > 0$ (or $\gamma_m^2 > 0$) that propagate in the $x$ direction, with an infinite number of eigenmodes with $\beta_m^2 < 0$ (or $\gamma_m^2 < 0$) that are evanescent in $x$. Similar expansion methods were used previously in Reference.35 for water waves and for quantum waveguides. In Eq. (1), we impose the outgoing boundary condition (no incoming wave outside the slab) to restrict our analysis to leaky guided resonances and bound states. Odd-in-$x$ TM modes can be written similarly by replacing the cosine in Eq. (1) with sine. $C_m$ and $T_m$ are coefficients of the eigenmode expansion, and they can be determined via the continuity of $E_z$ and $\partial E_z/\partial x$ at $x = 0.5h$, which requires

$$\sum_m C_m u_m(y) = \sum_m T_m \vartheta_m(y), \qquad (3.1)$$

$$-\sum_m i\gamma_m T_m \vartheta_m(y) = \sum_m C_m \beta_m \tan(0.5\beta_m h) u_m(y). \qquad (3.2)$$

The standing waves inside the slab [the $\cos(\beta_m x)$ in Eq. (1)] are superposition of waves propagating in $+x$ and in $-x$ directions, so one can interpret the in-slab fields as waves circulating within the slab with reflection and transmission at the two slab surfaces; the transverse phase shift for every propagating component is an integer multiple of $2\pi$ after a round trip with two reflections. This transversal resonance principle is common for guided modes in arbitrary waveguides.

While the two sets of eigenmodes $\{\vartheta_m\}$ and $\{u_m\}$ each form an infinite-dimensional basis, the high-order ones correspond to fast oscillating fields that are negligible at low frequencies. Therefore, in our numerical calculations, we will perform a truncation down to $M$ terms by expanding the eigenmodes in an $M$-dimensional basis (details below). In the truncated basis, the transformation between the two bases $\{\vartheta_m\}$ and $\{u_m\}$ is given by to an $M \times M$ matrix $\mathbf{P}$ such that $[u_1(y) \quad \ldots \quad u_M(y)] = [\vartheta_1(y) \quad \ldots \quad \vartheta_M(y)]\mathbf{P}$. In this way, Eq. (3) can be written in matrix form as

$$\boldsymbol{T} = \mathbf{P}\boldsymbol{C}, \tag{4.1}$$

$$-i\boldsymbol{\gamma}\boldsymbol{T} = \mathbf{P}\mathbf{B}\boldsymbol{C}, \tag{4.2}$$

where $\boldsymbol{T} = [T_1, \ldots, T_{M-1}, T_M]^T$, $\boldsymbol{C} = [C_1, \ldots, C_{M-1}, C_M]^T$, and $\boldsymbol{\gamma}$, $\mathbf{B}$ are diagonal matrices $\boldsymbol{\gamma} = \text{Diag}(\gamma_1, \ldots, \gamma_M)$, $\mathbf{B} = \text{Diag}(\beta_1 \tan(0.5\beta_1 h), \ldots, \beta_M \tan(0.5\beta_M h))$. Note that the matrix $\mathbf{B}$ is purely real since $\beta_m^2$ is real for all $m$. Eq. (4) is a linear equation group for vectors $\boldsymbol{T}$ and $\boldsymbol{C}$. Substituting Eq. (4.1) into Eq. (4.2) yields $(i\boldsymbol{\gamma}\mathbf{P} + \mathbf{P}\mathbf{B})\boldsymbol{C} = \mathbf{0}$. Non-trivial solutions exist when

$$f(k_y, \omega) \equiv \|i\boldsymbol{\gamma}\mathbf{P} + \mathbf{P}\mathbf{B}\| = 0, \tag{5}$$

where $\|*\|$ denotes the determinant of the matrix. Therefore, solving for Eq. (5) for a given $k_y$ yields the dispersion relation $\omega(k_y)$ for the bound states, resonances, as well as BICs. The corresponding field profiles are given by the vectors $\boldsymbol{T}$ and $\boldsymbol{C}$ through Eq. (1). At frequencies in the continuum spectrum of the surrounding medium ($\omega > k_y c/\sqrt{\varepsilon_B}$ for a homogeneous medium), some of the $\gamma_m$'s are real, and $f(k_y, \omega)$ is generally complex-valued; in such region finding the zeroes of $f(k_y, \omega)$ requires searching for solutions on the lower half of the complex-frequency plane with $\omega_r = \text{Re}(\omega)$ and $\omega_i = -\text{Im}(\omega)$ being the parameters. The imaginary part of the frequency is the decay rate, and the quality factor of the resonance is $Q = \omega_r/(2\omega_i)$.

The transformation matrix $\mathbf{P}$ is deduced from Eq. (2). Inspired by Ref. [24], we expand both sides of Eq. (2) in Fourier series and pick Fourier orders from $-N$ to $N$ (with a total of $M = 2N + 1$ terms). Then, Eq. (2) can be written as matrix equations

$$\mathbf{H}_1 \boldsymbol{\Phi} = \boldsymbol{\Phi}\boldsymbol{\beta}^2, \tag{6.1}$$

$$\mathbf{H}_2 \boldsymbol{\Theta} = \boldsymbol{\Theta}\boldsymbol{\gamma}^2, \tag{6.2}$$

with $\boldsymbol{\beta} = \text{Diag}(\beta_1, \ldots, \beta_M)$, and $\mathbf{H}_l$ is an $M \times M$ matrix whose $(m, m')$-th element is $-(2\pi n/a + k_y)^2 \delta_{nn'} + k_0^2 \xi_l(n - n')$, where $n = m - N - 1$, $n' = m' - N - 1$, and $\xi_l(n)$ is the $n$-th Fourier coefficient of $\varepsilon_l(y)$. The $m$-th column of $\boldsymbol{\Phi}$ (or $\boldsymbol{\Theta}$) contains the –N-to-N Fourier coefficients of $u_m$ (or $\vartheta_m$). $\mathbf{\Phi}$ and $\mathbf{\Theta}$ are transform matrices connecting $\{\vartheta_m\}$ and $\{u_m\}$ to the same basis with plane-wave elements, hence $\mathbf{P} = \mathbf{\Theta}^{-1}\mathbf{\Phi}$. Note that when the permittivity is real and mirror symmetric in $y$, [$\varepsilon_l(y) = \varepsilon_l^*(y) = \varepsilon_l(-y)$], the matrices $\mathbf{H}_l$ are real symmetric, so $\boldsymbol{\beta}^2$ and $\boldsymbol{\gamma}^2$ are real; moreover, $\mathbf{\Phi}$, $\mathbf{\Theta}$, and $\mathbf{P}$ can be chosen to be purely real.

**Solving dispersion equation for BIC**

BICs arise when the decay rate $\omega_i$ of a resonance becomes zero, or equivalently when the amplitudes of the radiating waves vanish: $T_{m\prime} = 0$ for all the $m\prime$ with $\gamma_{m\prime}^2 > 0$. Numerically it can be ambiguous to determine whether $\omega_i$ is very small or identically zero. Therefore, we use a slightly modified scheme to look for exact BICs. Define $f\prime$ as $f\prime = f|_{(\gamma_{m\prime}=0)}$: the propagation constants of the radiating waves $\gamma_{m\prime}$ are artificially set to zero; this function $f\prime$ is purely real for a lossless dielectric structure that is symmetric in $y$ (where $\mathbf{H}_l$ is real symmetric). A BIC not only satisfies $f = 0$; it also satisfies $f\prime = 0$ since $T_{m\prime} = 0$ for a BIC, and from Eq (3.2) it can be seen that setting $\gamma_{m\prime}$ to zero does not change the solution. Therefore, to search for BICs, we first solve the real-valued equation $f\prime(k_y, \omega) = 0$ at each $k_y$ for a real-valued frequency $\omega_i$; the solution also provides a mode profile given by $\boldsymbol{T}$ and $\boldsymbol{C}$. However, such a mode profile will only satisfy the continuity condition, Eq. (3), if $T_{m\prime} = 0$. In this work, we study the frequency range where there is only one leaky channel (only one $m\prime$ with $\gamma_{m\prime}^2 > 0$), and we perform a root finding with $k_y$ being the free parameter to look for solutions of $f\prime = 0$ where the amplitude of this radiation channel vanishes, $T_{m\prime} = 0$. Once found, such a solution will be a true bound state at a purely real frequency and with no radiation.

**Results and Discussion**

**Photonic crystal slab with homogeneous background**

In this work we study two systems. The first one is a layered slab in a homogeneous medium (Fig. 1b). It consists of a sequence of dielectric rectangles of size $d \times h$ with permittivity $\varepsilon_A$, surrounded by a homogeneous material with permittivity $\varepsilon_B$. The eigenmodes in the homogeneous medium are simply plane waves with $\gamma_{n+N+1} = \sqrt{\varepsilon_B k_0^2 - (k_y + 2\pi n/a)^2}$ and $\mathbf{\Theta}$ being an identity matrix. In **Fig. 2a**, the region in dispersion space with one leaky channel (one real $\gamma_{m\prime}$) is shaded in yellow. In the slab, the

Fourier coefficients of the permittivity is $\xi_1(n) = (d/a)(\varepsilon_A - \varepsilon_B)\text{sinc}(nd/a) + \varepsilon_B \delta_n$. For the basis truncation, we take $M = 21$ Fourier terms and eigenmodes ($N = 10$). As an example, we take $\varepsilon_A = 4.9$, $\varepsilon_B = 1$, $d = 0.5a$, and $h = 1.4a$. The band structure calculated by solving Eq. (5) is shown in Fig. 2a. For even-in-$x$ modes (blue curve), a non-symmetry-protected BIC occurs at $k_y = 0.3156\ (2\pi/a)$, $w = 0.4612\ (2\pi c/a)$. For odd-in-$x$ modes (green curve), a non-symmetry-protected BIC occurs at $k_y = 0.1640\ (2\pi/a)$, $w = 0.6006\ (2\pi c/a)$. Symmetry protected BICs can be found at the $\Gamma$ point ($k_y = 0$), where radiation vanishes because $E_z$ is odd in $y$ for the resonance but even in $y$ for the radiating wave. Fig. 2b shows the quality factor $Q = \omega_r/(2\omega_i)$ of the resonances; as expected, $Q$ diverges at the BICs. Fig. 2c shows the field profiles of the four BICs marked in Fig. 2a,b.

In Fig. 2d, we plot the coefficients $\boldsymbol{C}$ and $\boldsymbol{T}$ of the even-in-$x$ non-symmetry-protected BIC. Inside the slab (upper panel), the amplitudes $\boldsymbol{C}$ are dominated by two propagating modes (shown in red). Outside the slab (lower panel), there is only one propagating mode, and its amplitude vanishes at the BIC. The amplitude of the propagating mode outside the slab is the transmission from the in-slab modes, through Eq. (3). Therefore, the disappearance of radiation arises from destructive interference of the transmission from the in-slab modes, which, as shown in the upper panel, primarily consist of two propagating modes.

**Photonic crystal slab with periodically-modulated background**

The second system we consider is a PhC slab surrounded by a periodically-modulated background (Fig. 1c). The out-of-slab region has the same period as the slab but with a different fractional volume. We consider $\varepsilon_A = 4.9$, $\varepsilon_B = 1$, $d_1 = 0.5a$, $d_2 = 0.25a$, $h = 1.4a$. Fig. 3a indicates the region with one leaky channel ($\gamma_1^2 > 0$) using yellow shade, and plots the band structure obtained from Eq. (5); the corresponding quality factor is shown in Fig. 3b. In this structure, symmetry-protected BICs can be found both at the $\Gamma$ point ($k_y = 0$) and at the zone edge ($k_y = \pi/a$) inside the yellow-shaded region; symmetry incompatibility in $y$ prevents the in-slab mode from coupling to radiation, as can be seen from the corresponding mode profile in Fig. 3c.

The preceding examples concern structures where the dielectric is real and symmetric in $y$, for which the matrices $\mathbf{H}_l$ are real symmetric and so the function $f'$ is real-valued. However, BICs can exist in even more general systems. As long as $\varepsilon_l(-y) = \varepsilon_l^*(y)$, $\mathbf{H}_l$ is real (although not necessarily symmetric and not necessarily Hermitian). In such PT-symmetric systems, if the non-Hermiticity is below the PT-breaking threshold, the eigenvalues and eigenvectors of $\mathbf{H}_l$ can still be real, and the function $f'$ can still be real-valued. Such systems can also support BICs. However, if the introduction of gain leads to lasing, one will need to account for the nonlinearity resulting from gain saturation[36-38], which is beyond the linear model considered in this work.

**Conclusion**

We have presented a mode expansion method that can efficiently and quantitatively describe guided resonances and BICs in PhC slabs, and the method also reveals their underlying formation mechanisms. We find symmetry-protected BICs at the Γ point and at the zone edge, as well as BICs not protected by symmetry. The formalism is easily extendable and applicable to a wide range of structures. This is an attractive approach for the study of guided resonances and BICs in periodic structures.

**Acknowledgements**

This work was partly supported by the Army Research Office through the Institute for Soldier Nanotechnologies under contract no. W911NF-13-D-0001. B.Z. and M.S were partly supported (analysis and reading of the manuscript) by S3TEC, an Energy Frontier Research Center funded by the US Department of Energy under grant no. de-sc0001299. B.Z. was partially supported by the United States-Israel Binational Science Foundation (BSF) under award no. 2013508. C.W.H. was partly supported by the National Science Foundation through grant no. DMR-1307632. H. C., X. G. and X. L. were partly supported by the National Natural Science Foundation of China under Grants No. 61322501, No. 61574127, and No. 61275183, the Top-Notch Young Talents Program of China, the Program for New Century Excellent Talents (NCET-12-0489) in University, the Fundamental Research Funds for the Central Universities, and the Innovation Joint Research Center for Cyber-Physical-Society System.

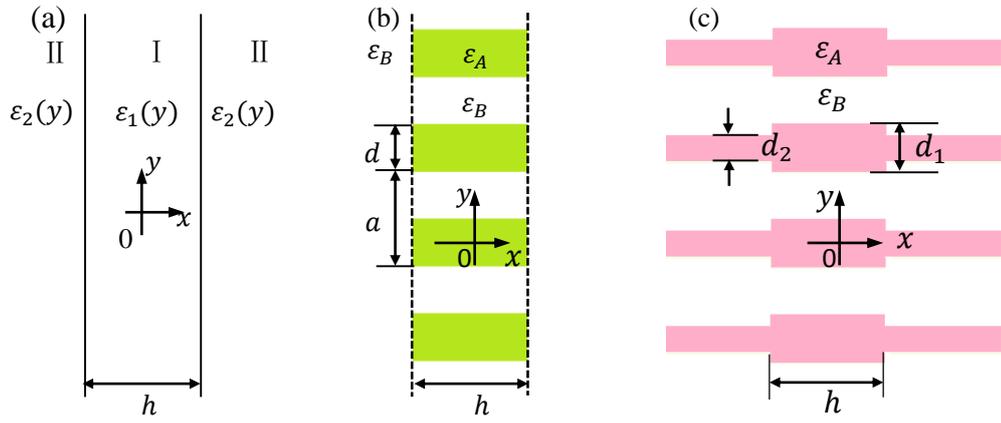

**Figure 1** (a) Structure: a 1D photonic crystal (PhC) slab with permittivity $\varepsilon_1(y)$ surrounded by another dielectric media with permittivity $\varepsilon_2(y)$. $\varepsilon_1$ and $\varepsilon_2$ have the same period of $a$, $\varepsilon_1(y+a) = \varepsilon_1(y)$, $\varepsilon_2(y+a) = \varepsilon_2(y)$. (b) Schematic of a 1D PhC slab embedded in homogeneous dielectric medium. (c) Schematic of a 1D PhC slab embedded in a periodic dielectric background.

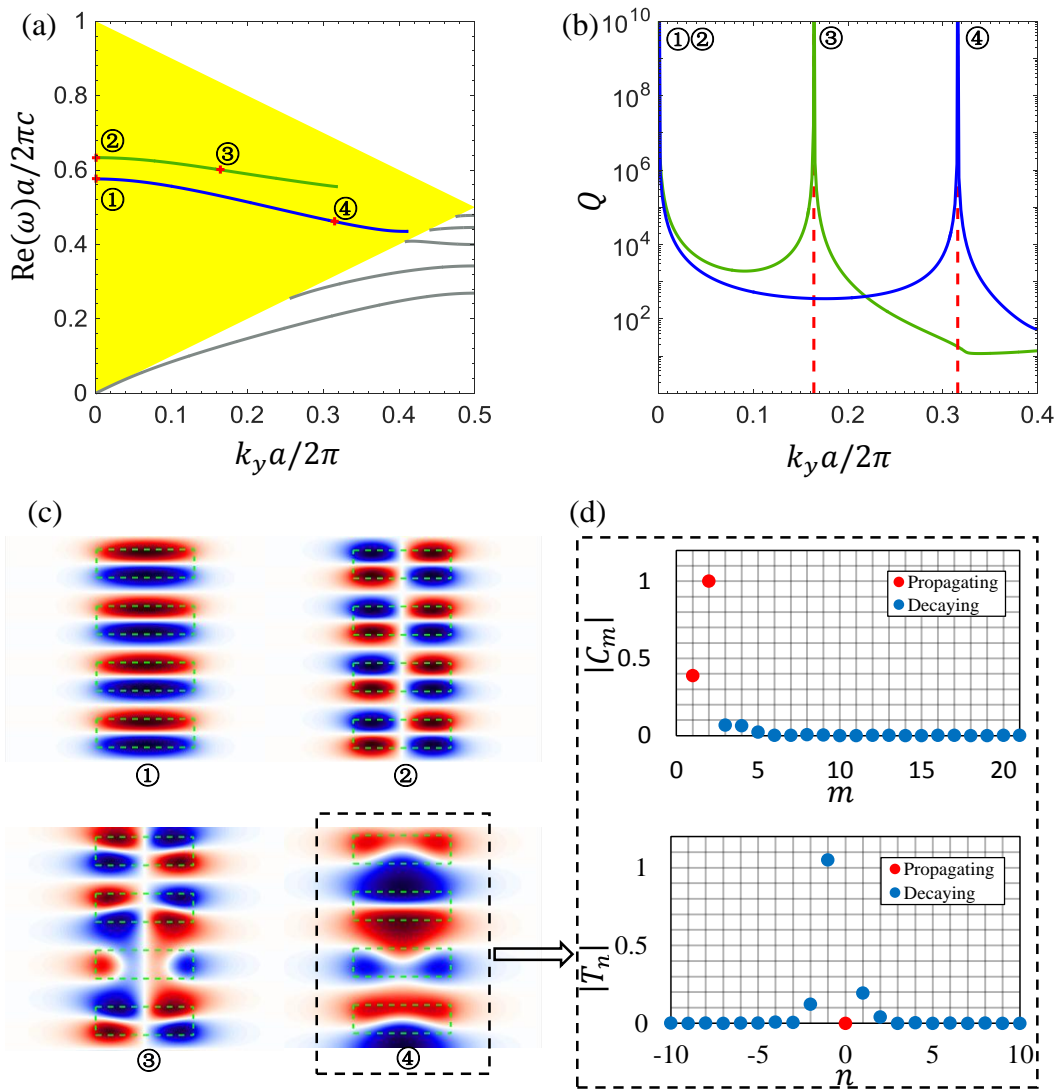

**Figure 2** (a) Band structure and (b) quality factors of the guided resonances in the system illustrated in Fig. 1b. (a) Yellow shaded area is where there is only one leaky channel in the surrounding medium. The blue (green) curve is the dispersion of guided resonances that are even (odd) in *x*. BICs are marked with red plus signs: ① and ② are protected by symmetry, while ③ and ④ are not. The grey curves are guided modes below the light line. Red dashed lines in (b) mark the location of BICs not protected by symmetry. (c) Electric field patterns of the BICs. (d) Magnitudes of the mode-expansion coefficients in the slab $C$ (upper panel) and outside the slab $T$ (lower panel) for the BIC ④. Red circles are components propagating in the *x* direction (where $\beta_m$ or $\gamma_m$ is real); blue circles are evanescent components with imaginary wave vector along the *x* direction. The structural parameters are: $\varepsilon_2 = 4.9$, $\varepsilon_1 = 1$, $h = 1.4a, d = 0.5a$.

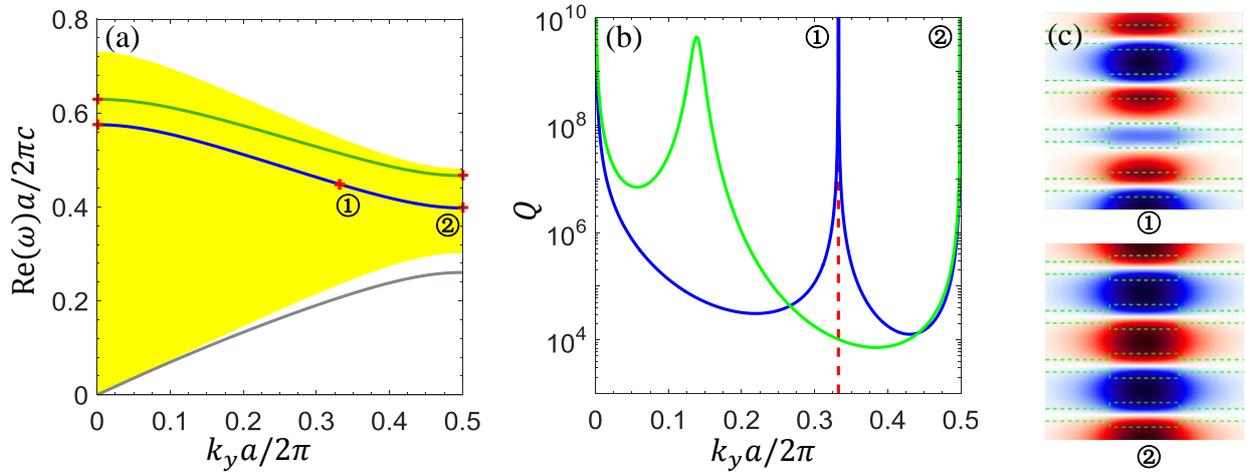

**Figure 3** (a) Band structure and (b) quality factors of the guided resonances in the system illustrated in Fig. 1c. The convention is the same as Fig. 2. A symmetry-protected BIC at the zone-edge is labeled by ②, and a BIC not protected by symmetry is labeled by ①. (c) Electric field patterns of the BICs ① and ②. The structural parameters are: $\varepsilon_A = 4.9$, $\varepsilon_B = 1$, $h = 1.4a, d_1 = 0.5a$, $d_2 = 0.25a$.